# Low-Frequency Noise Spectroscopy of Charge-Density-Wave Phase Transitions in Vertical Quasi-2D Devices

Ruben Salgado[1], Amirmahdi Mohammadzadeh[1], Fariborz Kargar[1], Adane Geremew[1], Chun-Yu Huang[1], Matthew A. Bloodgood[2], Sergey Rumyantsev[1,3], Tina T. Salguero[2] and Alexander A. Balandin[1,*]

[1]Nano-Device Laboratory (NDL) and Phonon Optimized Engineered Materials (POEM) Center, Department of Electrical and Computer Engineering, University of California, Riverside, California 92521 USA

[2]Department of Chemistry, University of Georgia, Athens, Georgia 30602 USA

[3]Center for Terahertz Research and Applications, Institute of High Pressure Physics, Polish Academy of Sciences, Warsaw 01-142 Poland

## Abstract

We report results regarding the electron transport in *vertical* quasi-2D layered 1T-TaS$_2$ charge-density-wave devices. The low-frequency noise spectroscopy was used as a tool to study changes in the *cross-plane* electrical characteristics of the quasi-2D material below room temperature. The noise spectral density revealed strong peaks – changing by more than an order-of-magnitude – at the temperatures closely matching the electrical resistance steps. Some of the noise peaks appeared below the temperature of the commensurate to nearly-commensurate charge-density-wave transition, possibly indicating the presence of the debated "hidden" phase transitions. These results confirm the potential of the noise spectroscopy for investigations of electron transport and phase transitions in novel materials.

[*]Corresponding author: balandin@ece.ucr.edu ; https://balandingroup.ucr.edu/

The charge-density-wave (CDW) phase is a macroscopic quantum state consisting of a periodic modulation of the electronic charge density accompanied by a periodic distortion of the atomic lattice in metallic crystals [1-3]. Recently, the field of CDW materials and devices experienced a true renaissance [4-16]. The renewed interest has been driven by layer-control of CDW materials, such as quasi-two-dimensional (2D) crystals of 1$T$-TaS$_2$ and other transition metal dichalcogenides (TMDs). Unlike classical bulk CDW materials with the quasi-1D crystalline structure, certain members of the layered TMD family exhibit unusually high transition temperatures to various CDW phases, opening up the possibility of practical applications for CDW devices [14-18]. The 2D crystal structure of TMDs allows one to exfoliate or grow layers with few-nanometer thicknesses, creating conditions for greater control of the CDW phase transitions with temperature or electric field, as well as enabling integration with other 2D materials [14-18].

One of the most interesting quasi-2D CDW TMD materials is 1$T$-TaS$_2$. At T=545 K, it undergoes a transition from the normal metallic phase to the incommensurate (IC) CDW phase; at T=355 K, it transforms to the nearly commensurate (NC) CDW phase; and finally, in the temperature range from T= 150 K to 180 K, it changes to the commensurate (C) CDW phase [4, 7, 8, 9, 13]. The NC-CDW phase consists of C-CDW domains separated by regions of IC-CDW phase. The high transition temperature of the IC to NC CDW phase, along with the possibility of controlling this transition with voltage, permits the practical implementation of such materials. Some of us recently demonstrated the 1$T$-TaS$_2$ voltage-controlled oscillator (VCO) based on the current switching driven by voltage-tuned CDW phase transitions [13]. These 1$T$-TaS$_2$ CDW devices, operational at room temperature (RT), exhibit radiation hardness [15], and can be used for information processing applications [17-18].

Almost all prior studies of CDW phenomena and CDW-based devices in 1$T$-TaS$_2$ and other TMDs have focused on the electron transport along the quasi-2D planes of these materials [4-19]. We are aware of only one prior detailed report on *cross-plane* electronic transport in vertical 1$T$-TaS$_2$ devices [19]. The cross-plane transport is expected to have interesting features because, in some layered TMDs, the crystal lattice distortion during the CDW phase transition affects the cross-plane direction even more strongly than the in-plane direction [12]. The previous study of the cross-plane transport revealed step-like changes in electrical resistivity at temperatures between 50 K and 100 K [19], which are substantially lower temperatures than those reported for the well-known transition between the C-CDW and NC-CDW transition [4, 7, 8, 9, 13]. These observations may be related to discussions about the possible existence and nature of metastable "hidden" states and phase transitions in 1*T*-TaS$_2$ [19-27]. A range of possible scenarios, *e.g.* Mott transitions or interlayer re-ordering of the stacking structure, are under consideration. Furthermore, there are indications that the "hidden" states can be particularly interesting in the vertical CDW devices [19].

Investigation of electron transport in the *vertical* quasi-2D 1*T*-TaS$_2$ CDW devices is more challenging than that in "conventional" planar CDW devices. The stress due to the mismatch of the temperature coefficients of the materials can affect the transport characteristics of the vertical devices more strongly than those of planar devices. In addition, it is often difficult to identify possible phase transitions in the relatively small changes in the current – voltage (I-V) characteristics. These considerations provide strong motivation for developing new experimental methods and approaches applicable to investigating cross-plane transport in layered CDW materials. Recently, we demonstrated that measurements of the low-frequency noise (LFN) can be used for identification of the transition between the nearly commensurate (NC) to incommensurate (IC) CDW phases and CDW sliding in a "conventional" planar 1*T*-TaS$_2$ device [14]. In this Letter, we show that LFN spectroscopy is effective for studying the electron transport and CDW transitions in the vertical 1*T*-TaS$_2$ devices, and in fact, this method can identify the CDW and potential "hidden state" transitions with higher accuracy than using the I-V characteristics.

High-quality 1*T*-TaS$_2$ crystals were prepared by the chemical vapor transport (CVT) method, from the elements using I$_2$ as the transport agent, via quenching from the crystal growth temperature. The details of this synthesis and corresponding material characterization data have been reported by some of us elsewhere [12, 13, 27]. The vertical CDW devices (see Figure 1 for schematic) were fabricated via mechanical exfoliation from CVT-grown crystals and an all-dry transfer method [26]. The device fabrication process can be described briefly in the following steps. First, the bottom electrodes were patterned by the electron beam lithography (EBL; LEO Supra) on a SiO$_2$/Si substrate. Immediately after, the layers of Ti/Au metal were deposited by the electron beam evaporation (EBE; Temescal BJD). The thin 1*T*-TaS$_2$ layers were exfoliated from bulk crystals onto an ultra-clean PDMS-glass stamp, which was mounted onto a home-built transfer stage [12, 13, 27 - 29]. The stage was equipped with a micromanipulator to align and transfer the exfoliated layers from the PDMS surface onto the bottom electrode. The thin layers of exfoliated hexagonal boron nitride (*h*-BN) were then placed overlapping the edge of 1*T*-TaS$_2$ layers using the same dry-stamp transfer technique in order to avoid any unwanted edge contacts. Finally, the top electrodes made of Ti and Au metals were fabricated using EBL and EBE.

[Figure 1: Device Schematic]

The optical images of two typical 1*T*-TaS$_2$ vertical CDW devices used in this study are shown in the inset to Figure 2 (a). The cross-sectional area and thickness of the 1*T*-TaS$_2$ layer in these devices were around 0.5 µm$^2$ and 90 nm, respectively. The temperature dependent current-voltage (I-V) characteristics of the devices were measured in a cryogenic probe station (Lakeshore TTPX) with a semiconductor analyzer (Agilent B1500). The devices were cooled down to 78 K at a ramp up rate of 1.5 K/min and heated at the same rate back to RT. The tested devices reached temperature stability between the heating and cooling cycle while their resistance was measured at DC bias sweep from 0 to 10 mV. In Figure 2 (a) we present the measured resistance of a representative two-terminal vertical 1*T*-TaS$_2$ device as a function of temperature. The resistance hysteresis associated with the commensurate (C) to nearly commensurate (NC) CDW phases can be observed at the transition temperature, $T_C$, which is in the range reported in previous studies [4, 7, 8, 9, 13]. During the measurements we observed changes in the resistance at temperatures below $T_C$. To investigate these changes, more 1*T*-TaS$_2$ vertical devices with similar structure and dimensions were fabricated and tested. Figure 2 (b) shows the temperature dependent resistance of three different vertical devices measured below the C-CDW – NC-CDW phase transition temperature. The measurements were performed in the cooling cycle. The resistance drop can be seen in all three measured devices, in the temperature range from 50 K to 100 K, which is substantially below the C-CDW – NC-CDW phase transition temperature.

[Figure 2: Resistance vs. Temperature]

The noise spectra were measured with a dynamic signal analyzer (Stanford Research 560) after the signal was amplified by the low-noise amplifier (Stanford Research 560). The devices under test (DUT) were DC biased with a "quiet" battery–potentiometer circuit in order to minimize the 60 Hz noise and its harmonics from the electrical grid. The noise measurements were conducted in the two-terminal device configuration. Because the contact resistance was negligibly small compared to the $1T$-TaS$_2$ layer resistance, the measured noise response was dominated by the CDW material of interest. The short-circuit current fluctuations were calculated following the conventional formula $S_I=S_V (R_L+R_D)/(R_L R_D)$, where $S_I$ is the current noise spectral density, $S_V$ is the voltage noise spectral density, $R_L$ and $R_D$ are the load and device resistances, respectively. Details of our low noise measurement procedures can be found elsewhere [14, 30 - 31].

Figure 3 (a) shows the voltage noise spectral density, $S_V$, as a function of frequency for a representative vertical $1T$-TaS$_2$ device under DC bias voltage, $V_B$, varying from 0.6 mV to 80 mV. The data were taken at RT. All spectra follow the $S_V \propto 1/f$ behavior without any signatures of generation-recombination (G-R) bulges. In Figure 3 (b) we present the current noise spectral density, $S_I$, as a function of the current though the device, which demonstrates perfect scaling with $I^2$. This proportionality implies that current does not drive the fluctuations but merely accentuates their visibility following Ohm's law [32]. It also suggests the absence of electromigration, strong self-heating or other damage to the device as a result of passing the current.

[Figure 3: Noise Characteristics]

We performed the low-temperature electrical resistance and LFN measurements on the vertical 1*T*-TaS$_2$ device as it was heated from 77 K to above the C-CDW – NC CDW phase transition temperature, T$_C$. In Figure 4, we present the measured electrical resistance and the normalized current noise spectral density, $S_I/I^2$, as the functions of temperature. The noise data were accumulated at $V_B$ = 13 mV and frequency of 10 Hz. The sudden drop in the resistance observed around 160 K is associated with the well-established C-CDW – NC-CDW phase transition. As one can see, the noise spectral density reveals a peak around the same temperature. The noise peak is clearly associated with this phase transition, in line with prior observation for the in-plane CDW devices [14]. The drop in resistance around T=100 K, observed in several vertical 1*T*-TaS$_2$ devices, is substantially below the C-CDW – NC-CDW phase transition temperature, and also is accompanied by the peak in the noise spectral density.

[Figure 4: Noise Spectroscopy]

It is interesting to note that the noise spectra measured at temperatures away from the phase transition temperature are always of the 1/*f* type (see also Ref. [14]). The noise spectra within the noise peak, which corresponds to the phase transition, have the form of the well-defined Lorentzian. Figure 5 shows an example of the noise spectra at temperature T=98 K, which corresponds to a proposed "hidden" phase transition, in the vertical configuration devices studied here. It is significant that the noise spectra measured at the same temperature but in "conventional" lateral CDW devices do not show signs of the Lorentzian shape. This fact suggests that the Lorentzian-type spectra exhibited in the vertical configuration is not a result of the G-R process but rather correspond to the phase transition.

[Figure 5: Noise at Different Temperature]

In conclusion, we have described the cross-plane electron transport in vertical quasi-2D layered 1*T*-TaS$_2$ CDW devices using low-frequency noise spectroscopy to study changes in the electrical characteristics below RT. We observed two steps in electrical resistivity in the temperature range from 150 K to 180 K, and in the range from 80 K to 85 K. The normalized low-frequency noise spectral density revealed strong peaks at these transition points, changing by more than an order-of-magnitude. The higher temperature feature was associated with the transition between the C-CDW and NC-CDW states. The lower temperature transition may indicate the presence of a proposed "hidden" CDW phase. These results further support the potential of the low-frequency noise spectroscopy for investigating electron transport in vertically-stacked quasi-2D materials.

**Acknowledgements**

The work at UCR and UGA was supported, in part, by the National Science Foundation (NSF) Emerging Frontiers of Research Initiative (EFRI) 2-DARE project: Novel Switching Phenomena in Atomic $MX_2$ Heterostructures for Multifunctional Applications (NSF EFRI-1433395). A.A.B. also acknowledges the UC - National Laboratory Collaborative Research and Training Program – University of California Research Initiatives (LFR-17-477237). S. R. also acknowledges partial support from the Center for Terahertz Research and Applications (CENTERA) in the framework of the International Research Agenda program of the Foundation for Polish Science co-financed by the European Union under the European Regional Development Fund. Nanofabrication was performed in the Center for Nanoscale Science and Engineering (CNSE) Nanofabrication Facility at UC Riverside.

**FIGURE CAPTIONS**

**Figure 1:** Schematic of the vertical quasi-2D 1*T*-TaS$_2$ charge-density-wave device. The electrical current flows perpendicular to the atomic planes of 1*T*-TaS$_2$ layered crystal. The direction of the current is shown with the red arrows.

**Figure 2:** (a) Electrical resistance of a representative vertical 1*T*-TaS$_2$ device measured in the cooling and heating cycles over a wide range of temperature. The insert shows an annotated optical microscopy image of the vertical device. (b) Electrical resistance of three different vertical 1*T*-TaS$_2$ devices measured below the well-known commensurate to nearly-commensurate charge-density-wave transition temperature. Note that all devices show steps in the resistance in the temperature range from 50 K to 100 K.

**Figure 3:** (a) Low-frequency voltage referred noise spectral density as a function of frequency for a vertical 1*T*-TaS$_2$ charge-density-wave device. The room-temperature data are shown for the bias voltage ranging from 0.6 mV to 80 mV. Note that all noise spectra are of the 1/f type without any signatures of the Lorentzian bulges. (b) Noise spectral density as a function of the electric current between two device terminals at a fixed frequency $f$=10 Hz.

**Figure 4:** Electrical resistance (upper panel) and normalized current noise spectral density (lower panel) as the functions of temperature. The noise data were measured at the bias of $V_{SD}$=13 mV and frequency $f$=10 Hz. The decrease in resistance at $T_C$=160 K corresponds to the well-known commensurate to nearly-commensurate CDW transition. A diagram, depicting the reconstruction of thirteen Ta atoms into hexagonal clusters, is shown to illustrate the phase transition. A distinctive noise peak is observed at the same temperature $T_C$. Below the C-CDW – NC-CDW phase transition temperature one can see another step in the resistance with the corresponding peak in the noise spectral density. The noise data are shown for two experimental runs to ensure reproducibility.

**Figure 5:** Normalized low-frequency current noise spectral density as a function of frequency. The data are presented for three different temperatures. The 1/*f* spectrum is shown for comparison with the dashed line. Note the Lorentzian peak in the noise spectrum taken new the transition temperature.

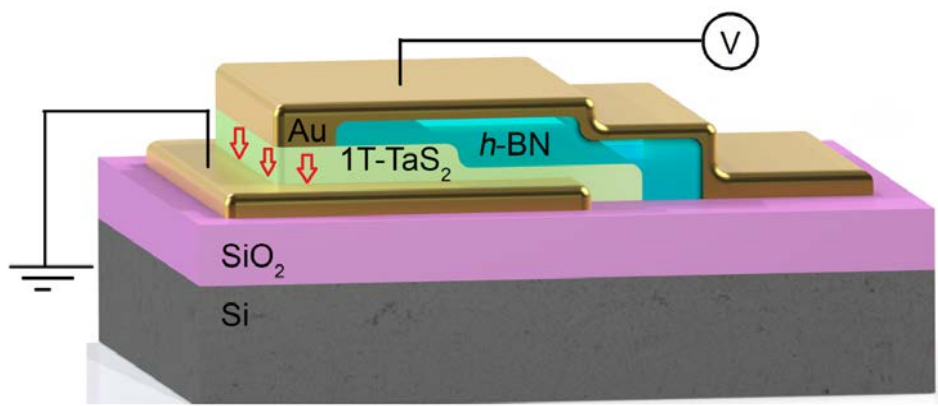

Figure 1

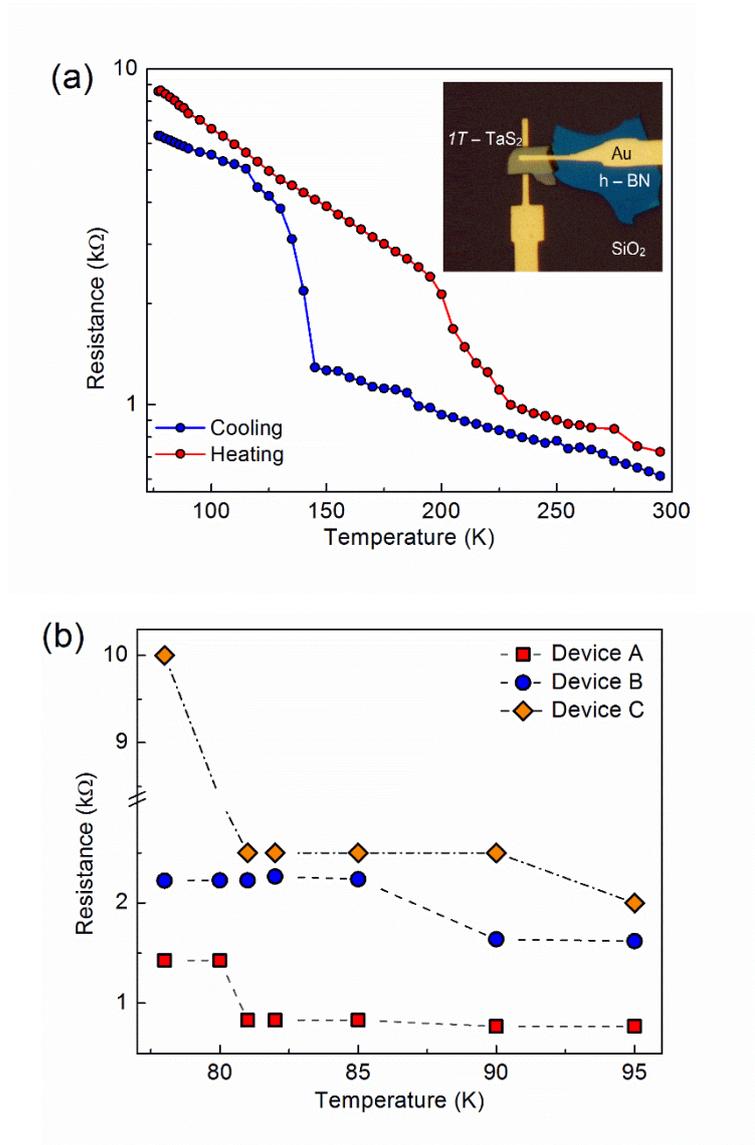

Figure 2 (a-b)

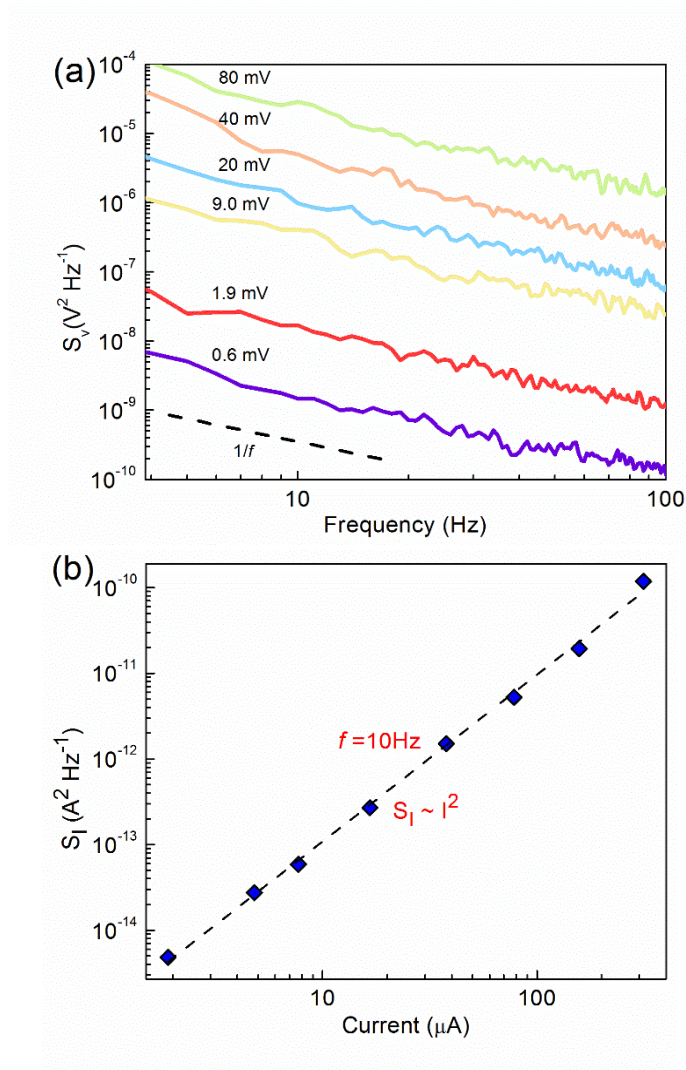

Figure 3 (a-b)

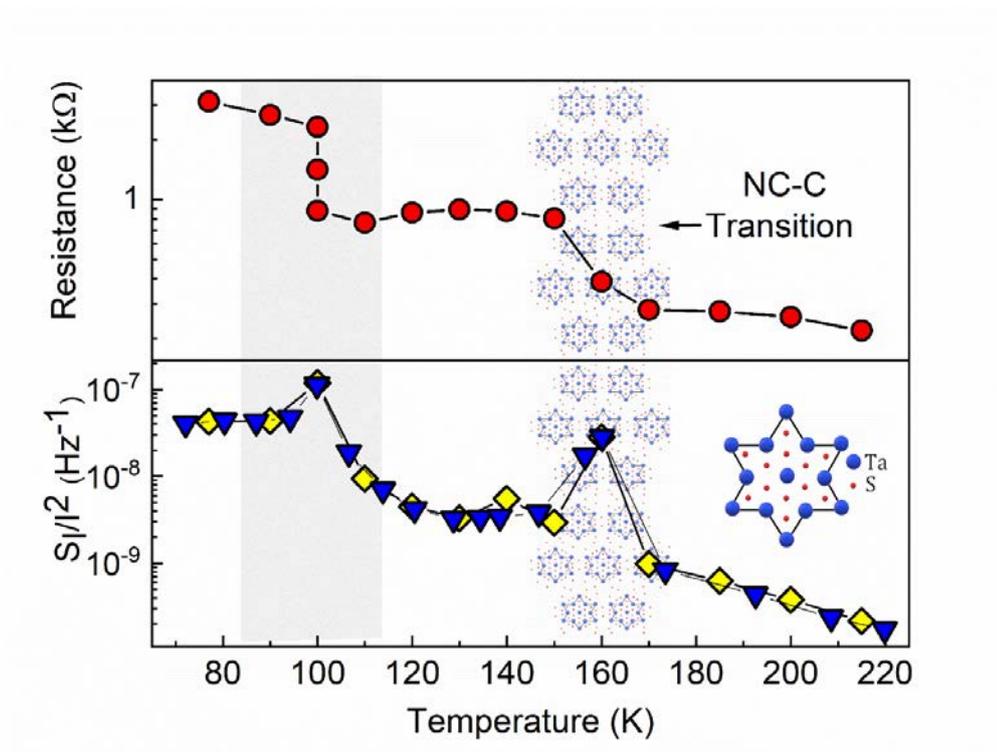

Figure 4

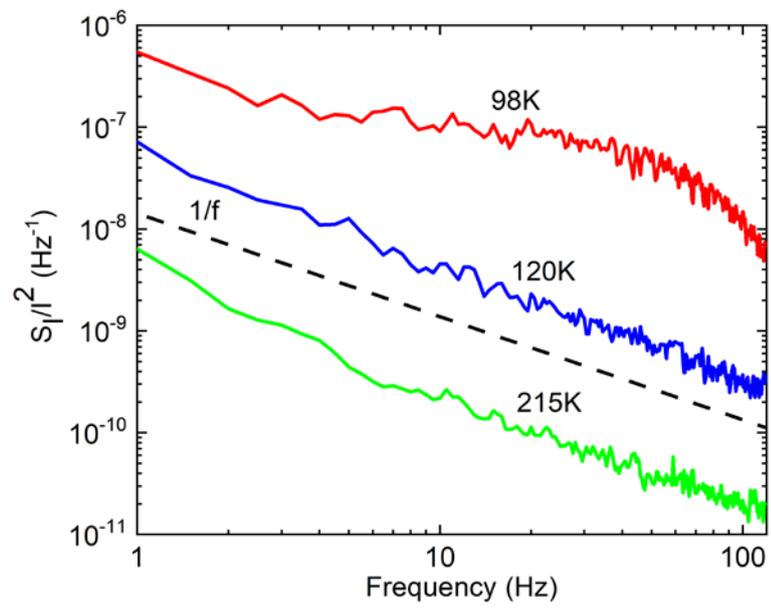

Figure 5